\documentclass[a4paper,fleqn]{cas-dc}

\usepackage[numbers]{natbib}
\usepackage{enumitem}
\usepackage{algorithm}
\usepackage{algpseudocode}
\usepackage{amsmath}
\usepackage{amsthm}

\theoremstyle{plain}
\newtheorem{definition}{Definition}[section]
\newtheorem{theorem}{Theorem}[section]
\newtheorem{corollary}[theorem]{Corollary}

\def\tsc#1{\csdef{#1}{\textsc{\lowercase{#1}}\xspace}}
\tsc{WGM}
\tsc{QE}
\tsc{EP}
\tsc{PMS}
\tsc{BEC}
\tsc{DE}

\begin{document}
\let\WriteBookmarks\relax
\def\floatpagepagefraction{1}
\def\textpagefraction{.001}
 \shorttitle{On Fixed-Parameter Tractability of Weighted 0-1 Timed Matching Problem on Temporal Graphs}
 \shortauthors{R. Kumar, B. Mazumdar, S. Mandal}

\title [mode = title] {On Fixed-Parameter Tractability of Weighted 0-1 Timed Matching Problem on Temporal Graphs}



\author[1]{Rinku Kumar}
\ead{phd2301101005@iiti.ac.in}


\affiliation[1]{organization={Department of Computer Science and Engineering, Indian Institute of Technology Indore},
                city={Indore},
                postcode={453552}, 
                state={Madhya Pradesh},
                country={India}}

\author[1]{Bodhisatwa Mazumdar}[orcid=0000-0003-1883-4639]
\ead{bodhisatwa@iiti.ac.in}


                
\author[2]{Subhrangsu Mandal}[orcid=0000-0002-0957-492X]
\cormark[1]
\fnmark[1]
\ead{santu.cst@gmail.com}


\affiliation[2]{organization={Department of Computer Science and Engineering, Indian Institute of Technology (ISM) Dhanbad},
                city={Dhanbad},
                postcode={826004}, 
                state={Jharkhand},
                country={India}} 







 \fntext[fn1]{S. Mandal was supported, in part, by FRS research grant MISC0100 provided by Indian Institute of Technology (ISM) Dhanbad.}


\begin{abstract}
Temporal graphs are introduced to model systems where the relationships among the entities of the system evolve over time. In this paper, we consider the temporal graphs where the edge set changes with time and all the changes are known a priori. The underlying graph of a temporal graph is a static graph consisting of all the vertices and edges that exist for at least one timestep in the temporal graph. The concept of 0-1 timed matching in temporal graphs was introduced by Mandal and Gupta [DAM2022] as an extension of the matching problem in static graphs. A 0-1 timed matching of a temporal graph is a non-overlapping subset of the edge set of that temporal graph. The problem of finding the maximum 0-1 timed matching is proved to be NP-complete on multiple classes of temporal graphs. We study the fixed-parameter tractability of the maximum 0-1 timed matching problem. We prove that the problem remains to be NP-complete even when the underlying static graph of the temporal graph has a bounded treewidth. Furthermore, we establish that the problem is W[1]-hard when parameterized by the solution size. Finally, we present a fixed-parameter tractable (FPT) algorithm to address the problem when the problem is parameterized by the maximum vertex degree and the treewidth of the underlying graph of the temporal graph.    
\end{abstract}



\begin{keywords}
Weighted $0-1$ Timed Matching \sep $0-1$ Timed Matching \sep Weighted Temporal Matching \sep Temporal Matching \sep Temporal Graphs \sep Dynamic Graphs \sep Fixed-Parameter Tractability 
\end{keywords}

\maketitle

\section{Introduction}

Graphs are an important tool to represent pairwise relationships between entities in a system. In many systems, the relationships between the entities and/or other properties in the system become temporal in nature. These systems result into networks with time-dependent topology. Examples of such networks include transportation networks \cite{lordan2020dynamic}, social networks \cite{iribarren2009impact}, biological networks \cite{lebre2010statistical}, communication networks \cite{fan2024temporal}, etc. Temporal graphs are introduced to model these networks, where the network topology changes with time. A temporal graph is a graph in which different properties of the graph change with time. In this paper, we consider the temporal graphs where only the edge set changes with time, and the changes in the graph topology are known a priori. These types of temporal graph can be represented by associating one or more non-overlapping time intervals on each edge in the temporal graph. These time intervals denote the time of existence of that particular edge. 


The added time dimension in a temporal graph introduces unique challenges to traditional graph structure related problems. In many cases, new definitions of the problems are required that incorporate the time dimension. Thus, problems of constructing different graph structures such as minimum spanning tree \cite{mandal2020convergecast}, dominating set \cite{mandal2018approximation}, colouring \cite{mertzios2021sliding}, etc. on temporal graphs have received considerable attention of researchers. Due to the dynamic nature of the graph topology, the matching problems become significantly more complex in temporal graphs. Different works \cite{baste2020temporal, mandal2022maximum, mertzios2023computing} in the literature have defined multiple variants of the matching problem on temporal graphs. The \textit{maximum 0-1 timed matching} problem on temporal graphs is introduced in \cite{mandal2022maximum} as an extension of the matching problem on static graphs. The problem of finding maximum 0-1 timed matching on temporal graphs is proved to be NP-complete \cite{mandal2022maximum} in multiple classes of temporal graphs, such as temporal tree, bounded degree temporal graphs, and bipartite temporal graphs. In this paper, we study the fixed-parameter tractability of the maximum weighted 0-1 timed matching problem. The key contributions of this paper are as follows.

\noindent {\em Our Contributions:} In this paper, we explore the fixed-parameter tractability of the maximum weighted 0-1 timed matching problem on temporal graphs. In particular, we have proved that this problem is NP-complete even when the underlying graph of the temporal graph has bounded treewidth. The problem is also proved to be $W[1]$-hard when parameterized by the solution size. Finally, a fixed-parameter tractable algorithm to address the maximum weighted 0-1 timed matching problem on temporal graphs is proposed when parameterized by the maximum vertex degree and the treewidth of the underlying graph of the given temporal graph.

The rest of this paper is organised as follows. Section \ref{sec:relwork} discusses some related works available in the literature. Section \ref{sec:sysmod} details the system model and assumptions about the temporal graphs considered in this paper. Section \ref{sec:defn} formally defines the problem and the terminology related to the temporal graphs. Section \ref{sec:maxweight} describes the details of the results. Section \ref{sec:conclude} concludes the paper.

\section{Related Work}
\label{sec:relwork}
The maximum 0-1 timed matching problem is an extension of the classical maximum matching problem on static graphs to the area of temporal graphs. The matching problem on static graphs is solvable in polynomial time. Edmonds has proposed the first algorithm~\cite{Edmonds1965} to solve the maximum matching problem in $O(n^4)$ time on a static graph with $n$ vertices. Due to its wide range of applications, the matching problem is studied in different classes of static graphs \cite{even1975n2,hopcroft1973n,mucha2006maximum} in the literature. 

The dynamic nature of the graph topology imposes additional challenges while addressing the matching problem in temporal graphs, where edges are active only at specific timesteps. Thus, multiple variants of the matching problem are introduced and studied on temporal graphs. In \cite{michail2016traveling}, Michail et al. have addressed the decision version of the temporal matching problem on a temporal graph where the objective is to check for a maximum cardinality matching $M$ on the underlying graph of the temporal graph such that each edge is assigned with a different timestep chosen from the timesteps when that edge exists. This problem is proved to be NP-hard. In \cite{baste2020temporal}, Baste et al. have defined a version of matching called $\gamma$-matching. An edge is a $\gamma$-edge if it exists for at least consecutive $\gamma$ timesteps in the temporal graph. The maximum $\gamma$ matching is the maximum cardinality subset of the $\gamma$ edges such that no two $\gamma$ edges are incident on a common vertex at any timestep. The problem is proved to be NP-complete when $\gamma > 1$ and a $2$-approximation algorithm is proposed to address the problem. They have proposed a kernelization based fixed-parameter tractable algorithm parameterized by the solution size to address the problem. In another work \cite{picavet2021temporal}, Picavet et al. have addressed the problem of $\gamma$ matching on geometric temporal graphs of bounded density. They have proposed a dynamic programming based exponential time exact algorithm and a PTAS to address the problem. Mertzois et al. have introduced $\Delta$-matching \cite{mertzios2023computing} on temporal graphs. A $\Delta$-matching $M$ can include two edge instances at the timesteps $t$ and $t'$ if those two do not share a common vertex or $|t-t'| > \Delta$ where $\Delta$ is the time window size. This problem is APX-hard when $\Delta > 2$ and the life time of the temporal graph is more than $3$. This problem is NP-hard even on temporal paths. They have proposed a $\frac{\Delta}{2\Delta - 1}$-factor approximation algorithm to address the problem. They have also proposed a fixed-parameter tractable algorithm parameterized by time window size and the size of maximum matching of the underlying graph to address the problem. In another work \cite{mandal2022maximum}, based on the concept of overlapping edges, Mandal et al. defined 0-1 timed matching on temporal graphs. Two edges are overlapping if they are incident on a common vertex and there exists at least one timestep $t$ when both the edges exist. The maximum 0-1 timed matching on a temporal graph is the maximum cardinality subset of edges such that no two edges in the subset are overlapping with each other. They have proved that this problem is NP-complete on temporal trees when two or more time intervals are associated with each edge. It is also proved that the problem is NP-complete on bounded degree bipartite temporal graphs even when each edge is associated with a single time interval. An $O(n\log n)$ time algorithm is proposed to address the problem on temporal tree with $n$ vertices when each edge is associated with a single time interval. An approximation algorithm is proposed to address the problem in general temporal graphs.      

Apart from these works, there are works \cite{bampis2018multistage,chimani2022approximating} in the literature that have addressed the multi-stage version of the matching problem on temporal graphs. Recently the study of fixed-parameter tractability of different graph problems on temporal graphs have received considerable attention of researchers. Different fixed-parameter tractable algorithms are proposed to address different problems in temporal graphs such as matching \cite{baste2020temporal,mertzios2023computing}, shortest path \cite{casteigts2021finding}, colouring \cite{marino2022coloring}, etc. To the best of our knowledge, there is no work in the literature that has studied the fixed-parameter tractability of the maximum 0-1 timed matching problem on temporal graphs.

\section{System Model}
\label{sec:sysmod}
In this paper, a temporal graph $\mathcal{G}$ is represented using the \textit{evolving graphs} \cite{ferreira2002models} model. In this model, a temporal graph is represented as a discrete finite sequence of static graphs. Each static graph at a certain timestep $t$ represents the state of the temporal graph at timestep $t$. The length of the sequence is referred to as the \textit{lifetime} of the temporal graph. Let $\mathcal{T}$ be the lifetime of $\mathcal{G}$, then $\mathcal{G}$ exists in the time interval $[0, \mathcal{T})$. We assume that the vertex set $\mathcal{V}$ remains unchanged throughout the lifetime of the temporal graph. Only the edge set $\mathcal{E}$ changes with time. We also assume that all changes in the edge set are known a priori. Additionally, it is assumed that there is no self-loop, and at most one edge can exist between any two vertices at a given timestep. Thus, the temporal graph $\mathcal{G = (V, E)}$ is represented as a sequence of static graphs, $(G_0, G_1, \cdots, G_{\mathcal{T} - 1})$ where each $G_i = (\mathcal{V}, \mathcal{E}_i)$ is the static graph at timestep $i$ with vertex set $\mathcal{V}$ and edge set $\mathcal{E}_i$ consisting of edges present at timestep $i$. As we assume that only the edge set changes over time, a temporal graph can be equivalently represented by assigning non-overlapping time intervals to each edge in the temporal graph. We also assume that each edge $e$ is assigned a non-negative real number representing its cost, $\omega (e)$, which remains unchanged throughout the lifetime of the temporal graph. Thus, an edge $e \in \mathcal{E}$ incident on two distinct vertices $u, v \in \mathcal{V}$ is  represented as $e(u, v, \omega(e), (s_1, f_1), (s_2, f_2), \cdots, (s_k, f_k))$, where $f_k \leq \mathcal{T}$. Each interval $(s_i, f_i)$ associated with an edge indicates that the edge exists for the time interval $[s_i,f_i)$, i.e., $e$ is present in $(G_{s_i}, G_{s_i + 1}, \cdots, G_{f_i - 1})$ where $0\le s_i < f_i \le \mathcal{T}$. When the temporal graph is unweighted, we represent the edge as $e(u, v, (s_1, f_1), (s_2, f_2), \cdots, (s_k, f_k))$. Since each interval associated with an edge is non-overlapping with any other interval associated with that edge, the maximum number of intervals associated with an edge is $\left\lfloor \frac{\mathcal{T}}{2}\right\rfloor$. An edge with a single time step is an instance of that edge. We denote an edge between vertices $u$ and $v$ as $e_{uv}$ when the exact time intervals associated with the edge are not required. The instance at timestep $t$ of an edge $e_{uv}$ is denoted as $e^t_{uv}$.      



\section{Preliminaries and Problem Definition}
\label{sec:defn}
In this section, we define the {\em maximum weighted 0-1 timed matching problem} on temporal graphs. We first define terminologies related to temporal graphs that are required to define the problem. Let $\mathcal{G} = (\mathcal{V}, \mathcal{E})$ be the temporal graph with lifetime $\mathcal{T}$, where $\mathcal{V}$ is the set of vertices and $\mathcal{E}$ is the set of edges.

\begin{definition}
\textbf{Underlying Graph:} The underlying graph of a temporal graph $\mathcal{G} = (\mathcal{V}, \mathcal{E})$, denoted as $\mathcal{G}_U = (\mathcal{V}, \mathcal{E}_U)$, is a static graph where $\mathcal{E}_U = \{(u, v) \mid \exists t \in [0, \mathcal{T}-1)$ such that $ e_{uv}^t$ is an instance of $e_{uv} \in \mathcal{E}\}$.
\end{definition}

\begin{figure*}[b!]
    \centering
    \includegraphics[width=0.6\linewidth]{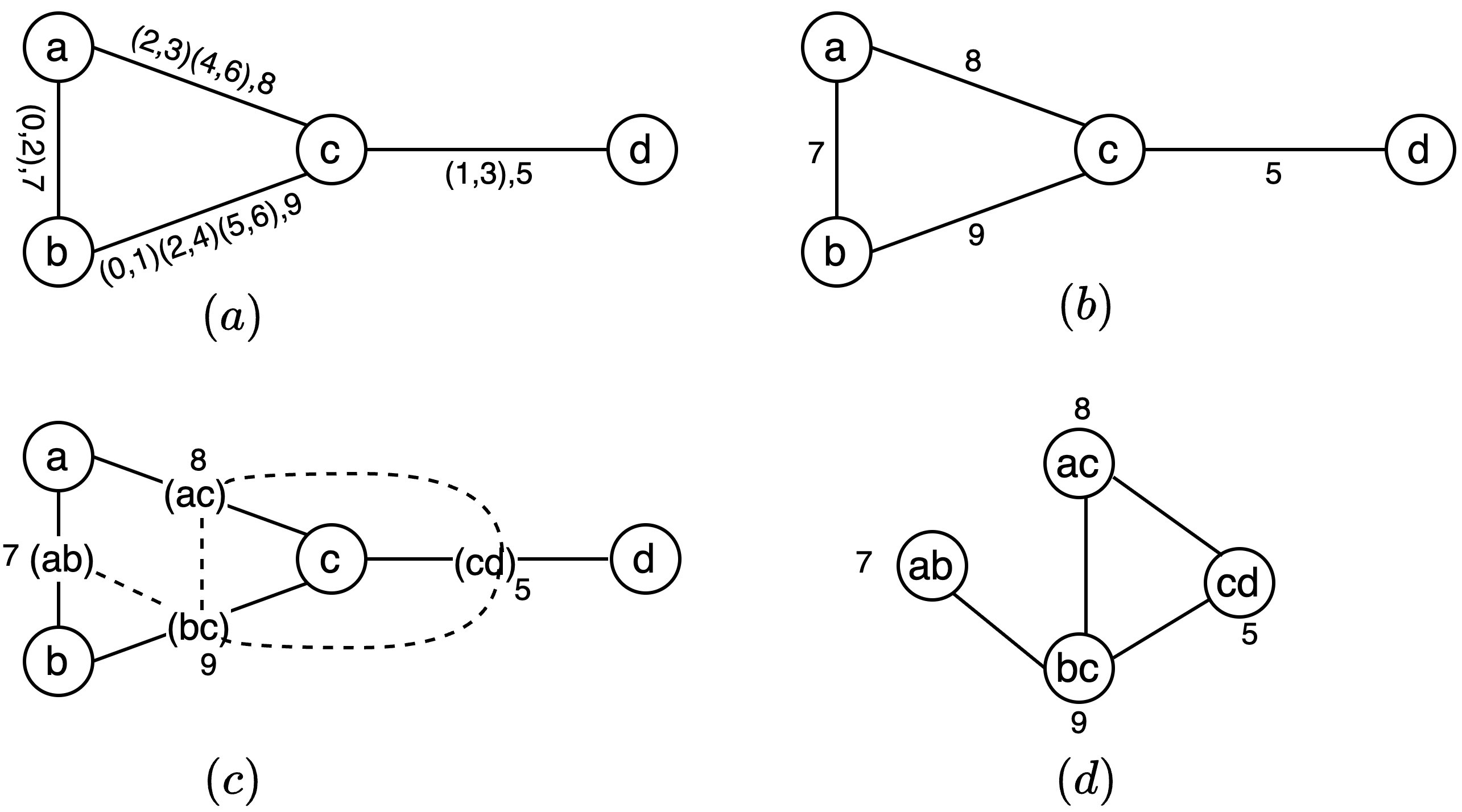}
    \caption{$(a)$ A temporal graph $\mathcal{G}$ with lifetime $6$, $(b)$ Underlying graph $\mathcal{G}_U$ of $\mathcal{G}$, $(c)$ Conversion to edge-overlap graph, $(d)$ The edge-overlap graph $\mathcal{G}_O$ of $\mathcal{G}$.} 
    \label{fig:temporal}
\end{figure*}

\begin{definition}
\textbf{Bounded Degree Temporal Graph:}  A temporal graph $\mathcal{G} = (\mathcal{V}, \mathcal{E})$ is a bounded degree temporal graph if the degree of each vertex $v \in \mathcal{V}$ in the underlying graph $\mathcal{G}_U$ is bounded by an integer $\Delta$, i.e., $deg(v)\leq \Delta$.
\end{definition}
Fig.~\ref{fig:temporal}(a) shows a bounded degree temporal graph $\mathcal{G}$ with $\mathcal{T} = 6$ where degree of each vertex in the underlying graph is bounded by $\Delta=3$. Fig.~\ref{fig:temporal}(b) shows the underlying graph of $\mathcal{G}$.

\begin{definition}
\textbf{Temporal Tree:}  A temporal graph $\mathcal{G} = (\mathcal{V}, \mathcal{E})$ is a temporal tree if the underlying graph $\mathcal{G}_U$ of $\mathcal{G}$ is a tree.
\end{definition}

\begin{definition}
\textbf{Overlapping Edges:} Two edges $e_{uv}, e_{vw} \in \mathcal{E}$ overlap each other if and only if both are incident at a vertex $v$ and there is at least one timestep $t$ when both the edges exist.
\end{definition}
In Fig.~\ref{fig:temporal}(a), edges $e_{ab}$ and $e_{ac}$ both are incident at the vertex $a$, but there is no timestep $t$ when both edges exist. Thus, these two are non-overlapping edges. If we consider edges $e_{ac}$ and $e_{cd}$, both edges exist at $t = 2$. Thus, these two are overlapping edges. 

\begin{definition}
\textbf{0-1 Timed Matching:} A 0-1 timed matching $M$ on a temporal graph, $\mathcal{G} = (\mathcal{V}, \mathcal{E})$ is a subset of $\mathcal{E}$ such that any two edges in $M$ are non-overlapping with each other.
\end{definition}

\begin{definition}
\textbf{Maximum Weighted 0-1 Timed Matching:} A maximum weighted 0-1 timed matching on a temporal graph $\mathcal{G} = (\mathcal{V}, \mathcal{E})$ is a 0-1 timed matching, $\mathcal{M} \subseteq \mathcal{E}$ such that the sum of weights of the edges in $\mathcal{M}$ are maximum among all such possible 0-1 timed matching on $\mathcal{G}$. 
\end{definition}
In the temporal graph shown in Fig.~\ref{fig:temporal}(a), the maximum weighted 0-1 timed matching $\mathcal{M}=\{e_{ab}, e_{ac}\}$ with total weight $15$. Next, we introduce the concept of {\em edge-overlap graph} of a temporal graph that we are going to use while addressing this problem.

\begin{definition}
\textbf{Edge-Overlap Graph:} The edge-overlap graph of a temporal graph $\mathcal{G} = (\mathcal{V}, \mathcal{E})$ is a static graph, $\mathcal{G}_O= (V_O,E_O)$, where the vertex set $V_O$ includes a vertex $v_{uv}$ for each edge $e_{uv} \in \mathcal{E}$ and the edge set $E_O$ includes the edge $(v_{uv}, v_{wx})$ connecting $v_{uv}, v_{wx} \in V_O$ when $e_{uv}, e_{wx}$ are overlapping with each other. 
\end{definition}
Note that when $\mathcal{G(V, E)}$ is an edge-weighted temporal graph, such that each edge $e_{uv} \in \mathcal{E}$ is associated with weight $\omega{(e_{uv})}$, then the edge-overlap graph $\mathcal{G}_O (V_O, E_O)$ is a vertex-weighted static graph where weight $\omega{(e_{uv})}$ is assigned to the vertex $v_{uv} \in V_O$. Fig.~\ref{fig:temporal}(d) shows the edge-overlap graph $\mathcal{G}_O$ of the edge-weighted temporal graph $\mathcal{G}$ shown in Fig~\ref{fig:temporal}(a).

\section{Maximum Weighted 0-1 Timed Matching Problem}
\label{sec:maxweight}
The unweighted maximum 0-1 timed matching problem is proved to be NP-complete \cite{mandal2022maximum} on different restricted classes of temporal graphs, such as temporal tree, bounded degree bipartite temporal graphs. In this section, we study the fixed-parameter tractability of the {\em maximum weighted 0-1 timed matching} problem on temporal graphs. In particular, we prove that the unweighted maximum 0-1 timed matching problem is NP-complete on bounded treewidth temporal graphs, and this problem is W[1]-hard when parameterized by the solution size. Then, we propose an algorithm for temporal graphs when the underlying graph of the temporal graph has bounded treewidth and bounded degree. 
In Theorem 5.1 of \cite{mandal2022maximum}, it is proved that the maximum 0-1 timed matching problem is NP-complete on temporal tree. We get the following result from this theorem.

\begin{corollary}
    \label{cor:treewidth} The maximum 0-1 timed matching problem is NP-complete even when the underlying graph of the temporal graph has bounded treewidth.  
\end{corollary}

In Theorem 6.5 of \cite{mandal2022maximum}, it is proved that there is an approximation preserving reduction from {\em maximum independent set} problem to the maximum 0-1 timed matching problem. In this paper, we use the same idea to show a parameterized reduction from the independent set problem to the 0-1 timed matching problem. The parameterized version of the independent set problem is defined as: 

\begin{itemize}[leftmargin=*]
    \item[] {\bf Input:} A graph $G (V, E)$ and an integer $k$.
    \item[] {\bf Question:} Is there a subset $I \subset V$ of exactly size $k$ such that for any $u, v \in I$, $(u,v) \notin E$?
\end{itemize}

The parameterized version of the 0-1 timed matching problem is:

\begin{itemize}[leftmargin=*]
    \item[] {\bf Input:} A temporal graph $\mathcal{G (V, E)}$ and an integer $k'$.
    \item[] {\bf Question:} Is there a subset $M \subset \mathcal{E}$ of exactly size $k'$ such that no two edges $e_{uv}, e_{wx} \in M$ are overlapping with each other?
\end{itemize}

Next, we prove that there is a parameterized reduction from the independent set problem to the 0-1 timed matching problem.

\begin{theorem}
    \label{thm:pared}
    There exists a parameterized reduction from the independent set problem to the 0-1 timed matching problem.
\end{theorem}

\begin{proof}
    Consider an instance of instance $(G, k)$ of the {\em independent set} problem where $G=(V, E)$ is a static graph with $|V|=n_V$, $|E| = m_E$ and $k$ is a given integer. We construct an instance $(\mathcal{G}, k')$ of the 0-1 timed matching problem. Let $S$ be a set of distinct integers, $S=\{0, 1, \ldots, m_E-1\}$. Each edge $(u, v)\in E$ is assigned with a distinct integer, $n_{uv}\in S$. Let $V_{0}\subseteq V$ be the set of zero-degree vertices in $V$ such that $|V_{0}|=n_{0}$. The set $V_0$ can be computed from $G$ in $O(n_V)$ time. Each vertex, $u \in V_0$ is assigned with a distinct integer, $n_u$ from the set of integers, $S'=\{m_E,\ldots, m_E+n_0-1\}$. For each vertex $v\in V$, the {\em neighborhood} of $v$ is defined as, $N_v=\{u\in V|(u, v)\in E\}$. Thus, $\forall u\in V_0, N_u=\phi$. We construct a temporal graph, $\mathcal{G}=(\mathcal{V}, \mathcal{E})$, as follows.
    \begin{enumerate}[leftmargin=*]
        \item[(i)] Initialize $\mathcal{V}=\phi, \mathcal{E}=\phi, \mathcal{E}_1=\phi, \mathcal{E}_2=\phi$. 
        \item[(ii)] $\forall v \in V$, we add a vertex $\mathsf{v}_v$ to $\mathcal{V}$, and one additional vertex $\mathsf{v}'$, i.e., $\mathcal{V} = \mathcal{V} \cup \{\mathsf{v}_v|\forall v \in V\} \cup \{\mathsf{v}'\}$.
        
        \item[(iii)] $\forall v \in V_0$, we add an edge between $\mathsf{v}_v$ and $\mathsf{v'}$. The time interval assigned to this edge is $(n_v, n_v+1)$, where $n_v\in S'$. So, $\mathcal{E}_1 = \mathcal{E}_1 \cup \{e (\mathsf{v}_v\mathsf{v}', (n_v, n_v+1))| \forall v\in V_0 \}$.
        
        \item[(iv)] $\forall v\in V\setminus V_0$, we add an edge between the corresponding $\mathsf{v}_v$ and $\mathsf{v}'$ in $\mathcal{V}$. The time intervals assigned to this edge are $(n_{vu_1}, n_{vu_1}+1), (n_{vu_2}, n_{vu_2}+1)\ldots, (n_{vu_{deg(v)}}, n_{vu_{deg(v)}}+1)$, where $u_1, u_2,\ldots, u_{deg(v)}\in N_v$. 
        So, $\mathcal{E}_2 = \mathcal{E}_2 \cup \{e (\mathsf{v}_v,\mathsf{v}',(n_{vu_1}, n_{vu_1}+1),\dots, (n_{vu_{\deg(v)}}, n_{vu_{\deg(v)}}+1))| \forall v\in V \setminus V_0 \}$.
        \item[(v)] The set of edges in the temporal graph,  $\mathcal{G}$ is formed as  $\mathcal{E}=\mathcal{E}_1 \cup \mathcal{E}_2$, and the lifetime of $\mathcal{G}$ is $m_E+n_0$.
        \item[(vi)] We assign $k' = k$.
    \end{enumerate}
    The constructed temporal graph $\mathcal{G}$ contains an edge between each vertex $v\in \mathcal{V}\setminus \{\mathsf{v}'\}$ and $\mathsf{v}'$. Moreover, there is no edge between any other pair of vertices in $\mathcal{G}$. Thus, $\mathcal{G}$ is a temporal tree. 
    
    For a given $0$-$1$ timed matching $M$ in $\mathcal{G}$, we construct an independent set $IS_{max}$ on $G$, where $|IS_{max}|=|M|$. We select the set of vertices $IS_{max}\subseteq V$, such that $IS_{max}=\{v|e_{\mathsf{v}_v\mathsf{v}'}\in M\}$. We show that $IS_{max}$ is an independent set in $G$ and $|IS_{max}|=|M|$. Now, for each edge in $M$, a vertex in $IS_{max}$ is selected, so $|IS_{max}|=|M|$. We prove by contradiction that $IS_{max}$ is an independent set for $G$. Suppose $IS_{max}$ is not an independent set of $G$. If so, then there are at least two vertices, $v_1, v_2\in IS_{max}$, such that $(v_1, v_2)\in E$. As $v_1, v_2\in IS_{max}$, then both $e_{\mathsf{v}_{v_1} \mathsf{v}'}$,  $e_{\mathsf{v}_{v_2} \mathsf{v}'} \in M$. Now, as $(v_1, v_2)\in E$, $(v_1, v_2)$ is assigned with integer $n_{v_1v_2}$. This implies that the time interval, $(n_{v_1v_2}, n_{v_1v_2}+1)$ is assigned to both $e_{\mathsf{v}_{v_1} \mathsf{v}'}$ and  $e_{\mathsf{v}_{v_2} \mathsf{v}'}$, and both these edges are incident on $\mathsf{v}'$. Hence, $M$ is not a $0$-$1$ timed matching on the temporal graph $\mathcal{G}$. This results in a contradiction, and hence $|IS_{max}|$ is an independent set for $G$.

    Next, for a given independent set $IS_{max}$ in $G$, we construct a $0$-$1$ timed matching $M$ in $\mathcal{G}$ such that $|IS_{max}|=|M|$. We construct a $0$-$1$ timed matching $M$ as $M = \{e_{\mathsf{v}_v\mathsf{v}'} | v \in IS_{max}\}$. Since we include an edge $e_{\mathsf{v}_v\mathsf{v}'}$ in $M$ for every $v$ in $IS_{max}$, $|M| = |IS_{max}|$. We prove that $M$ is a $0$-$1$ timed matching in $\mathcal{G}$. Assume that $M$ is not a $0$-$1$ timed matching and two overlapping edges $e_{\mathsf{v}_v\mathsf{v}'},e_{\mathsf{v}_u\mathsf{v}'} \in M$. As $e_{\mathsf{v}_v\mathsf{v}'},e_{\mathsf{v}_u\mathsf{v}'} \in M$, both $u,v \in IS_{max}$. The edges $e_{\mathsf{v}_v\mathsf{v}'},e_{\mathsf{v}_u\mathsf{v}'}$ are overlapping means there is a timestep $t$ when both the edges exist. From the construction of $\mathcal{G}$, this implies that $t$ is an integer that is assigned to an edge which is incident on both $u,v$. This implies that $IS_{max}$ is not an independent set in $G$. Hence, $M$ is a $0$-$1$ timed matching in $\mathcal{G}$. This completes the proof.  
\end{proof}

It is already known that the maximum independent set problem is $W[1]$-hard \cite{cygan2015parameterized} when parameterized by the solution size. From this fact and Theorem \ref{thm:pared} we get the following theorem.

\begin{theorem}
    The maximum $0$-$1$ timed matching problem is $W[1]$-hard when parameterized by the solution size.
\end{theorem}

Next, we prove that there is a fixed-parameter tractable algorithm that solves the weighted $0$-$1$ timed matching problem on temporal graphs when parameterized by treewidth and maximum vertex degree of the underlying graph of the temporal graph. In particular, we transform the problem into a {\em maximum weighted independent set} (MWIS) on the edge-overlap graph of given temporal graph. We prove that the problem of finding the maximum weighted 0-1 timed matching problem on a temporal graph $\mathcal{G}(\mathcal{V,E})$ can be addressed by solving the MWIS problem on the edge-overlap graph $\mathcal{G}_O$ of the temporal graph $\mathcal{G}$. 

\begin{theorem}
 The problem of finding a maximum weighted 0-1 timed matching on a temporal graph $\mathcal{G}$ is solvable by finding a maximum weighted independent set on $\mathcal{G}_O$. 
 \label{thm:maxind}
\end{theorem}
\begin{proof} At first, we construct the edge-overlap graph $\mathcal{G}_O = (V_O, E_O)$ of the given temporal graph $\mathcal{G}$. Then, we construct a solution for finding a maximum weighted 0-1 timed matching on $\mathcal{G}$ from a solution to the problem of finding a maximum weighted independent set on the edge-overlap graph $\mathcal{G}_O$ of $\mathcal{G}$. We construct the edge-overlap graph $\mathcal{G}_O = (V_O, E_O)$ of a given temporal graph $\mathcal{G}(\mathcal{V,E})$ as follows.
    \begin{itemize}[leftmargin=*]
       \item For each weighted edge, $e_{uv} \in \mathcal{E}$, we include a weighted vertex $v_{uv}$ in $V_O$, such that the weight of the vertex $v_{uv}$ is $w(v_{uv})$ = $\omega(e_{uv})$. Thus,
       $V_O = \{v_{uv} \,|\, \forall e_{uv} \in \mathcal{E}\}$.
        
       \item The edge set $E_O$ of $G_O$ includes edges between two vertices $v_{uv}, v_{vw} \in V_O$ if $e_{uv}$ and $e_{vw}$ are overlapping with each other in $\mathcal{G}$. Thus, 
       $E_O = \{(v_{uv}, v_{vw}) \,|\, e_{uv}, e_{vw} \in \mathcal{E}$ are overlapping with each other$\}$.
       
    \end{itemize}
Since the determination of two overlapping edges can be done in polynomial time, the construction of edge-overlap graph of a given temporal graph is done in polynomial time. Next, we prove that if there is a solution for the MWIS problem on the edge-overlap graph then we can find a solution for the maximum 0-1 timed matching problem on the original temporal graph in polynomial time. 

Let $I \subseteq V_O$ be a maximum weighted independent set with total weight $w(I)$ in the edge-overlap graph $\mathcal{G}_O$ of the given temporal graph $\mathcal{G(V, E)}$. We construct a maximum weighted 0-1 timed matching $\mathcal{M}$ for $\mathcal{G}$ as follows. For each vertex, $v_{uv} \in I$, we add an edge $e_{uv} \in \mathcal{E}$ to $\mathcal{M}$. We prove that $\mathcal{M}$ is a maximum weighted 0-1 timed matching in $\mathcal{G}$.

\begin{itemize}
    \item At first, we prove that $\mathcal{M} \subseteq \mathcal{E}$. Since each vertex in $V_O$ is added for an edge in $\mathcal{E}$ and $I \subseteq V_O$, $\mathcal{M} \subseteq \mathcal{E}$. 
    \item Next, we prove that all the edges in $\mathcal{M}$ are non-overlapping with each other. We prove this by contradiction. Let there be two edges $e_{uv}, e_{vw} \in \mathcal{M}$ that overlap with each other. This implies that corresponding two vertices, $v_{uv}, v_{vw} \in V_O$ are included in $I$. Since $e_{uv}, e_{vw} \in \mathcal{M}$ are overlapping with each other, there is an edge between $v_{uv}, v_{vw} \in V_O$ from the construction of $\mathcal{G}_O$ from $\mathcal{G}$. Hence, it contradicts that $I$ is an independent set on $\mathcal{G}_O$. This proves that no two edges in $\mathcal{M}$ overlap with each other.
    \item We prove that $\mathcal{M}$ is a maximum weighted 0-1 timed matching on $\mathcal{G}$. We prove this by contradiction. Let $\mathcal{M}'$ be a $0$-$1$ timed matching, such that the total weight $\omega(\mathcal{M}')$ of the edges in $\mathcal{M}'$ is more than the total weight $\omega(\mathcal{M})$ of the edges in $\mathcal{M}$. We construct a set of vertices $I' \subseteq V_O$ by choosing the vertices in $V_O$ corresponding to the edges in $\mathcal{M}'$. For each edge, $e_{uv} \in \mathcal{E}$, there is a vertex $v_{uv} \in V_O$, such that $\omega(e_{uv}) = w(v_{uv})$. Thus, $w(I') = \omega(M')$ and $w(I) = \omega(\mathcal{M})$. Since $\omega(\mathcal{M}') > \omega(\mathcal{M})$, $w(I') > w(I)$. As $\mathcal{M}'$ is a 0-1 timed matching on $\mathcal{G}$, no two edges in $\mathcal{M}'$ overlap with each other. Thus, from the construction of $\mathcal{G}_O$, there are no edges between two vertices in $I'$. Hence $I'$ is an independent set for $\mathcal{G}_O$ with total weight $w(I') > w(I)$. This contradicts our assumption that $I$ is a maximum weighted independent set on $\mathcal{G}_O$. Thus, $\mathcal{M}$ is a maximum weighted 0-1 timed matching on $\mathcal{G}$. This completes the proof of this theorem.  \qedhere  
\end{itemize}
\end{proof}

Theorem \ref{thm:maxind} proves that if we have a solution to the problem of finding a maximum weighted independent set problem on the edge-overlap graph of a given temporal graph, then we can find a maximum weighted 0-1 timed matching for the given temporal graph in polynomial time. The problem of finding a maximum weighted independent set on a static graph is known to be NP-complete \cite{garey2002}. There is a fixed-parameter tractable algorithm \cite{bodlaender2013} to find a maximum weighted independent set on a static graph when parameterized by treewidth. In particular, there is an $O(2^kn)$ algorithm \cite{cygan2015parameterized} to find the maximum weighted independent set on a graph with treewidth $k$ and $n$ vertices. Next, we find the bound on the treewidth of the edge-overlap graph when the treewidth of the underlying graph of the temporal graph is $k$. To compute the bound, we use the known results on the treewidth of line graphs \cite{harvey2014treewidth}. It is known that for any static graph $G$ with treewidth $k$, the treewidth, $tw(L(G))$ of its line graph $L(G)$ is $tw(L(G)) \leq (k + 1)\Delta(G) - 1$, where $\Delta(G)$ is the maximum degree of a vertex in $G$. We extend this result in \cite{harvey2014treewidth} to find the bound on the treewidth of an edge-overlap graph of a temporal graph.     

\begin{theorem}
    Let $\mathcal{G}$ be a temporal graph and $\mathcal{G}_U$ be its underlying graph. If treewidth of $\mathcal{G}_U$ is $k$ and the maximum degree of a vertex in $\mathcal{G}_U$ is $\Delta(\mathcal{G}_U)$ then treewidth $tw(\mathcal{G}_O)$ of edge-overlap graph $\mathcal{G}_O(V_O, E_O)$ of $\mathcal{G}$ is $tw(\mathcal{G}_O) \leq (k + 1)\Delta(\mathcal{G}_U) - 1$. 
    \label{thm:egdetree}
\end{theorem}

\begin{proof}    
    In the line graph of $L(\mathcal{G}_U)$ of $\mathcal{G}_U$, every edge $e_{uv} \in \mathcal{G}_U$ is represented as a vertex $v_{uv} \in V_O$. Two vertices $v_{uv}, v_{vw} \in L(\mathcal{G}_U)$ are connected by an edge when $e_{uv}, e_{vw}$ are incident on the same vertex. In the edge-overlap graph $\mathcal{G}_O$, the vertex set is the same as $L(\mathcal{G}_U)$, only the edge set $E_O$ includes edges connecting two vertices $v_{uv}, v_{vw} \in V_O$ if these edges are overlapping with each other. Thus, $\mathcal{G}_O$ is a subgraph of $L(\mathcal{G}_U)$. Hence, we get the result $tw(\mathcal{G}_O) \leq (k + 1)\Delta(\mathcal{G}_U) - 1$.
\end{proof}

\subsection{Proposed Algorithm}
Using the results proved in Theorem \ref{thm:maxind} and \ref{thm:egdetree}, we propose a fixed parameter tractable algorithm to address the problem of finding maximum weighted 0-1 timed matching on a edge weighted temporal graph parameterized by the maximum vertex degree and treewidth of the underlying graph of that temporal graph. The algorithm starts by constructing the edge-overlap graph $G_O(V_O,E_O)$ for the input graph $\mathcal{G}(\mathcal{V},\mathcal{E})$. As the underlying graph $\mathcal{G}_U$ of the input temporal graph $\mathcal{G}$ has bounded max-degree and treewidth, the resulting $G_O(V_O, E_O)$ also has bounded treewidth. Next, a nice tree decomposition $(T,\{X_t\})$ of $G_O(V_O,E_O)$ is created using the algorithm presented in \cite{9719727}. In this tree decomposition, we apply the dynamic programming based algorithm described in \cite{cygan2015parameterized} to find the maximum weighted independent set $I$ on the edge overlap graph $G_O$. For each vertex in $I$, we select the corresponding edge in $\mathcal{G}$ to obtain the maximum weighted $0$-$1$ timed matching on $\mathcal{G}$.

\subsubsection{Time Complexity Analysis}

The construction of the edge-overlap graph, $G_O = (V_O, E_O)$, requires iterating over all vertices of the temporal graph, $\mathcal{G(V, E)}$, and checking the overlaps between the incident edges. For each vertex $v \in \mathcal{V}$, $O(\Delta^2)$ pairs of incident edges are checked, where $\Delta$ is the maximum degree of a vertex in the underlying graph $\mathcal{G}_U$ of $\mathcal{G}$. The check for overlapping edges for a pair of edges takes $O(\mathcal{T})$ time. Thus, the total cost for the construction of the edge overlap graph, $G_O$ is $O(|\mathcal{V}| \Delta^2 \mathcal{T})$,  where $|\mathcal{V}|$ is the number of vertices in $\mathcal{G}$ and $\mathcal{T}$ is the lifetime of $\mathcal{G}$.
Computing the tree decomposition of $G_O$ using the algorithm in \cite{9719727} that provides a tree decomposition of treewidth, $k' \le 2k+1$ takes $O(2^{O(k)} |V_O|)$ time, where $k$ is the treewidth of $G_O$ with $k\leq \frac{1}{2} (tw(\mathcal{G}_U) + 1) \Delta - 1$, and $tw(\mathcal{G}_U)$ is the treewidth of the underlying graph of $\mathcal{G}$. On this tree decomposition $(T, \{X_t\})$ of $G_O$, the dynamic programming algorithm to compute the maximum weighted independent set takes $O(2^{k'}(k')^2|V_O|)$ time. Thus, the total time complexity of the algorithm is $O(|\mathcal{V}| \Delta^2 \mathcal{T} + 2^{O(k)} |\mathcal{E}_U| + 2^{k'} (k')^2|\mathcal{E}_U|)$.

\section{Conclusion}
\label{sec:conclude}
We study the maximum weighted 0–1 timed matching problem in temporal graphs from the perspective of parameterized complexity. We prove that the problem remains NP-complete even when the underlying static graph of the temporal graph has bounded treewidth, demonstrating that structural sparsity alone does not yield tractability. Moreover, we show that the problem is W[1]-hard when parameterized by the size of the solution, indicating that this natural parameterization is unlikely to admit an FPT algorithm. We present a fixed-parameter tractable algorithm for the problem when the underlying static graph has both bounded treewidth and bounded maximum degree identifying a tractable regime within this otherwise hard problem space.

\bibliographystyle{cas-model2-names}

\bibliography{References}





\end{document}